\documentclass{article}

 \usepackage[main, final]{neurips_2025}

\usepackage[utf8]{inputenc} 
\usepackage[T1]{fontenc}    
\usepackage{hyperref}       
\usepackage{url}            
\usepackage{booktabs}       
\usepackage{amsfonts}       
\usepackage{nicefrac}       
\usepackage{microtype}      
\usepackage{xcolor}         
\usepackage{graphicx}
\usepackage{amsmath}

\usepackage{caption} 

\newcommand{\keywords}[1]{\textbf{Keywords:} #1}  

\title{
Label Uncertainty for Ultrasound Segmentation 
}

\author{
Malini Shivaram$^{1}$\And
Gautam Rajendrakumar Gare$^{1}$\And
Laura Hutchins$^{2}$ \And
Jacob Duplantis$^{2}$ \And
Thomas Deiss$^{2}$ \And
Thales Nogueira Gomes$^{2}$ \And
Thong Tran$^{2}$ \And
Keyur H. Patel$^{2}$ \And
Thomas H Fox$^{2}$ \And
Amita Krishnan$^{2}$ \And
Deva Ramanan$^{1}$ \And
Bennett DeBoisblanc$^{2}$ \And
Ricardo Rodriguez$^{3}$ \And
John Galeotti$^{1}$ \And
$^{1}$ Carnegie Mellon University, Pittsburgh, USA \And
$^{2}$ LSUHSC Internal Medicine, New Orleans, USA \And 
$^{3}$ Cosmetic Surgery Facility LLC, Baltimore MD 21093, USA \And
}

\begin{document}

\maketitle              
%


\begin{figure}[h!]
\centering
\begin{tabular}{@{}c@{\hspace{0.5em}}c@{\hspace{0.5em}}c@{\hspace{0.5em}}c@{\hspace{0.5em}}c@{\hspace{0.5em}}c@{\hspace{0.5em}}c@{\hspace{0.5em}}c@{\hspace{0.5em}}c@{}}
    LUS Image & Soft Label & $> 0$ & $\geq 20$ & $\geq 40$ & $\geq 50$ & $\geq 60$ & $\geq 80$ & $= 100$ \\
    \includegraphics[width=0.1\textwidth]{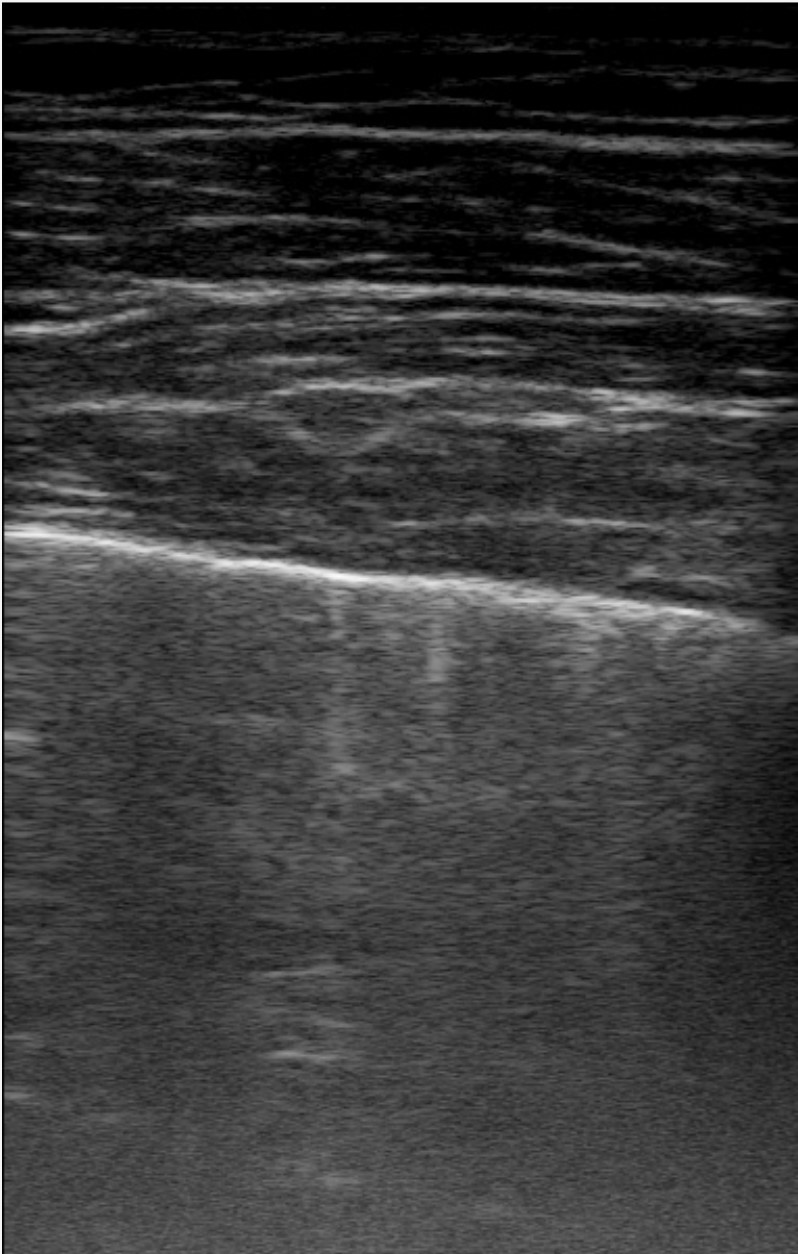} &

    \includegraphics[width=0.1\textwidth]{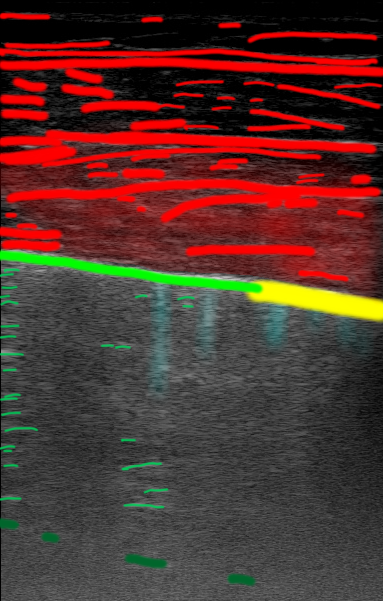} &
    \includegraphics[width=0.1\textwidth]{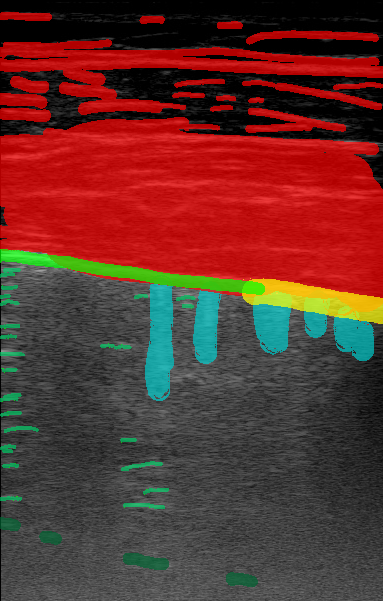} &
    \includegraphics[width=0.1\textwidth]{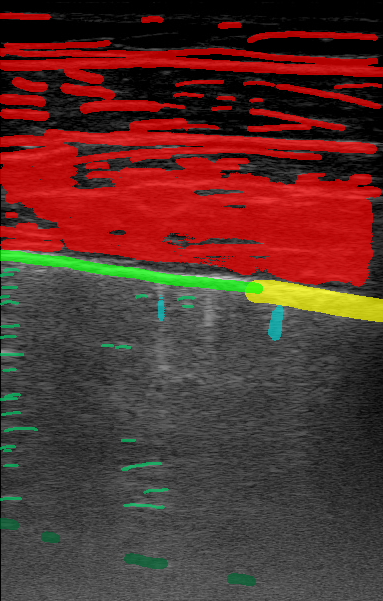} &
    \includegraphics[width=0.1\textwidth]{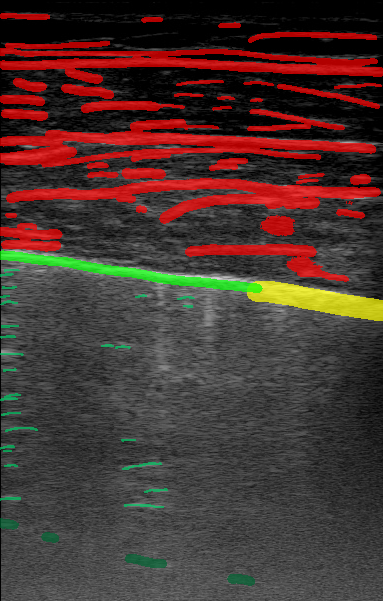} &
    \includegraphics[width=0.1\textwidth]{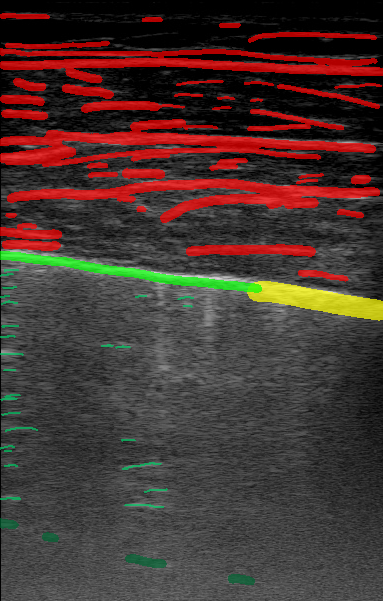} &
    \includegraphics[width=0.1\textwidth]{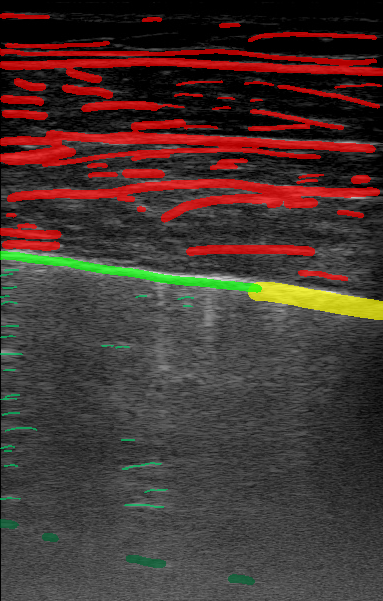} &
    \includegraphics[width=0.1\textwidth]{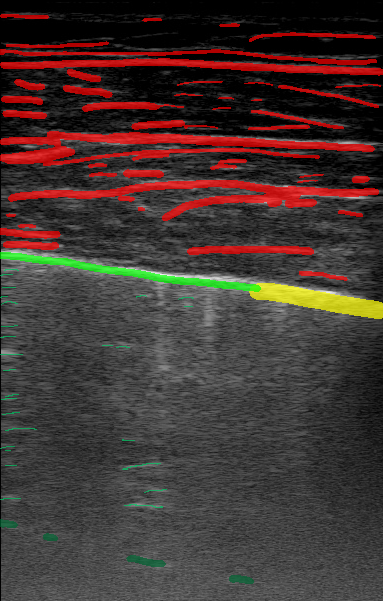} &
    \includegraphics[width=0.1\textwidth]{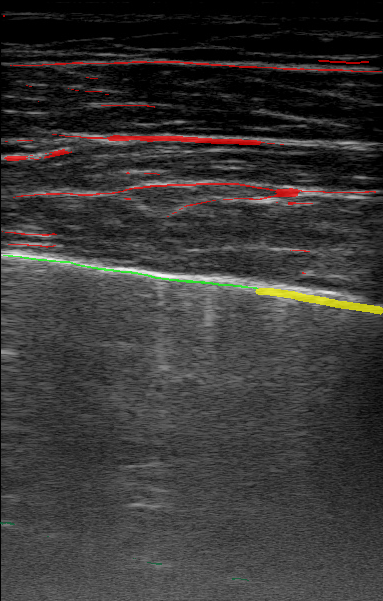}
\end{tabular}
\caption{Expert confidence labels thresholded at various confidence values for lung ultrasound (LUS) images.
We demonstrate that incorporating confidence values during training improves segmentation performance, which in turn enhances outcomes on clinically-critical downstream tasks. 
} 
\label{fig:confidence-label}
\end{figure}

\begin{abstract}
In medical imaging, inter-observer variability among radiologists often introduces 
label uncertainty, particularly in modalities where visual interpretation is subjective. Lung ultrasound (LUS) is a prime example—it frequently presents a mixture of highly ambiguous regions and clearly discernible structures, making consistent annotation challenging even for experienced clinicians. In this work, we introduce a novel approach to both labeling and training AI models using expert-supplied, per-pixel confidence values. Rather than treating annotations as absolute ground truth, we design a data annotation protocol that captures the confidence that radiologists have in each labeled region, modeling the inherent aleatoric uncertainty present in real-world clinical data. We demonstrate that incorporating these confidence values during training leads to improved segmentation performance. More importantly, we show that this enhanced segmentation quality translates into better performance on downstream clinically-critical tasks—specifically, estimating S/F oxygenation ratio values, classifying S/F ratio change, and predicting 30-day patient readmission. While we empirically evaluate many methods for exposing the uncertainty to the learning model, we find that a simple approach that trains a model on binarized labels obtained with a (60\%) confidence threshold works well. Importantly, high thresholds work far better than a naive approach of a 50\% threshold, indicating that training on {\em very} confident pixels is far more effective. Our study systematically investigates the impact of training with varying confidence thresholds, comparing not only segmentation metrics but also downstream clinical outcomes. These results suggest that label confidence is a valuable signal that, when properly leveraged, can significantly enhance the reliability and clinical utility of AI in medical imaging.




\end{abstract}

\section{Introduction}

Lung ultrasound (LUS) is an increasingly valuable diagnostic tool, particularly in point-of-care and resource-constrained clinical settings. Its non-invasive nature, portability, and real-time imaging capabilities make it ideal for assessing respiratory conditions, including pulmonary edema, pleural effusion, and pneumonia. However, the utility of LUS in machine learning applications, such as semantic segmentation and disease prediction, is limited by the inherent variability in expert annotations. Lung ultrasound is extra challenging to annotate due to poor image quality, artifacts, and attenuation of the ultrasound waves deeper in the body. These properties make it difficult to recognize various clinical structures and accurately determine their boundaries, even for expert clinicians. Different radiologists may interpret the same ultrasound frame differently, leading to label noise and uncertainty in supervised learning models.



Traditional segmentation datasets assume a single ground truth label per image, ignoring the variability and confidence associated with each annotation. Furthermore, as often in the case of radiological images, segmentation annotations from clinicians vary; two clinicians would likely produce different but equally correct segmentations. This can hinder model performance, particularly in clinical tasks where nuanced visual features drive decision-making, by causing unstable training or overfitting to a particular clinician's annotation style. To address this gap, we introduce the first LUS video dataset annotated with confidence scores for segmentation labels, capturing inherent radiologist aleatoric certainty at the pixel level.

In this study, we investigate how incorporating label confidence into the training pipeline affects model performance. We hypothesize that treating label uncertainty as a first-class signal,  instead of  noise, can improve both segmentation accuracy and downstream clinical predictions. Specifically, we evaluate segmentation models trained using varying confidence thresholds and assess their ability to predict S/F ratio change, S/F ratio values, and 30-day patient readmission, which are key metrics for healthcare outcomes. Our findings reveal that models trained with a 60\% confidence threshold outperform those trained with conventional labels, demonstrating the importance of accounting for annotation confidence in clinical machine learning.
\subsection{Related Work}

Traditional medical image segmentation relies on binary ground truth labels, which fail to capture the ambiguity present in real-world clinical imaging - particularly in cases with fuzzy anatomical boundaries or inter-observer disagreement. To address this, recent work has embraced soft labeling and uncertainty-aware techniques.

A foundational concept in this space is trimap segmentation \cite{trimap}, which partitions an image into definite foreground, background, and uncertain regions. While initially requiring manual annotation, automated approaches have evolved using statistical models and user-guided optimization. Trimaps are central to image matting, where they guide attention toward ambiguous regions to facilitate the estimation of per-pixel alpha masks that encode transparency or confidence values. However, conventional trimap-based methods often treat the uncertain region as a uniform class, disregarding internal gradations when modeling appearance for segmentation tasks. In contrast, we show that incorporating graded uncertainty within these regions can serve as a valuable learning signal, provided that their uncertainty is properly calibrated.



Extending this to medical domains, \cite{Wang2021MedicalUncertainty} proposed a model that outputs binary masks, alpha mattes, and uncertainty maps to better delineate ambiguous structures. Similarly, Confidence Contours \cite{Ye2023ConfidenceSegmentation} provide annotators with tools to explicitly mark both high- and low-confidence boundaries, improving both model interpretability and labeling efficiency.

Soft labeling also addresses supervision noise in datasets. For example, LF-Net \cite{Li2024LabelAnnotations} combines high-quality annotations with soft labels derived from annotator disagreement and structural priors. Soft supervision has further proven effective in domain-specific tasks like breast and thyroid cancer diagnosis \cite{Wang2022ADiagnosis}.

Uncertainty-aware evaluation metrics are also emerging. \cite{Stutz2025EvaluatingTruth} propose statistical aggregation to reflect ambiguous ground truths, while SoftSeg \cite{Gros2021SoftSeg:Segmentation} improves both segmentation calibration and accuracy through soft-label training.

Various modeling techniques - such as Bayesian neural networks \cite{ArbelADebates}, deep ensembles, and adversarial plausibility estimation - capture uncertainty probabilistically \cite{uncertainty-in-medical-imaging}. However, they often lack the ability to incorporate explicit, per-pixel confidence annotations provided by human experts. To address this limitation, we introduce a soft-brush annotation interface inspired by trimaps, which allows for continuous-valued (graded) confidence labeling. This approach enables efficient, single-expert confidence reporting and prioritizes downstream clinical utility over segmentation accuracy alone.

\section{Methodology}
We aim to segment relevant lung features on ultrasound images guided by confidence labels from human-labeled segmentations. In particular, we segment the pleural line (as sharp and fuzzy pleura), fascia bands, A lines, sub-A lines, and vertical lines (B-lines). To evaluate the clinical significance of our segmentations, we use them for three downstream tasks: classifying S/F ratio change between ultrasound video pairs, predicting a patient's S/F ratio, and predicting 30-day hospital readmission. These demonstrate the segmentation model's ability to perform in various settings that require different methods of leveraging and aggregating ultrasound data.

\begin{figure}[h!]
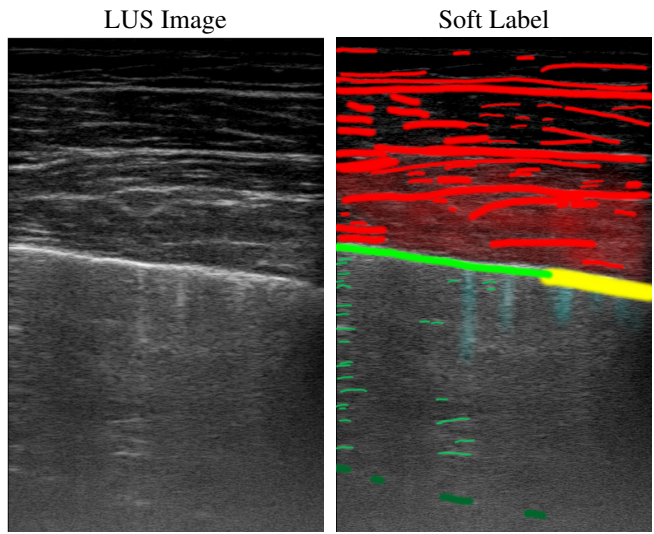

\centering
\begin{tabular}{@{}c@{\hspace{0.5em}}c@{\hspace{0.5em}}c@{\hspace{0.5em}}c@{\hspace{0.5em}}c@{\hspace{0.5em}}c@{\hspace{0.5em}}c@{\hspace{0.5em}}c@{\hspace{0.5em}}c@{}}
    LUS Image & Soft Label \\
    \includegraphics[width=0.3\textwidth]{U5a1_image.jpg} &

    \includegraphics[width=0.3\textwidth]{U5a1-rawlabel.png} &
\end{tabular}
\caption{Soft Confidence labeled image by an expert using the paint brush annotation tool to set the confidence as the transparency of the layer, along with corresponding lung ultrasound (LUS) image.
} 
\label{fig:soft-confidence-label}
\end{figure}

\subsection{Dataset}


We use an in-house lung ultrasound dataset 
of linear probe videos consisting of 189 patients (718 videos) with multiple (minimum 2 per patient) ultrasound B-scans of left and right lung regions at depths ranging from 4cm to 6cm under different scan settings, obtained using a Sonosite X-Porte ultrasound machine 
(IRB-approval no ****).
In particular, there are 6 scanning locations of the left and right lung regions, called \textit{views}. The R1 (Right 1) and L1 (Left 1) views (zone 1) correspond to ultrasound scans on the upper front part of the right and left chest, respectively. The R2 (Right 2) and L2 (Left 2) views (zone 2) are taken from the side of the chest, on the right and left. Finally, the R3 (Right 3) and L3 (Left 3) views (zone 3) capture images from the lower back or side areas of the right and left lungs. Ultrasound scans from different views allow for a comprehensive visual evaluation of lung regions.

In this study, we extract datapoints pertaining to only a single disease: Congestive Heart Failure (CHF), which results in a dataset containing a total of 42 patients. Focusing on lung ultrasounds for CHF patients is particularly valuable because CHF commonly leads to the accumulation of excess fluid in the lungs that creates recognizable patterns in lung ultrasound images. Furthermore, the severity of these lung ultrasound findings, particularly the number and distribution of B-lines, has been shown to correlate with important clinical metrics \cite{lus-chf}, including measures of oxygenation and risk for hospital readmission \cite{lus-chf-readmission}.

Our segmentation dataset consists of the first frame of each LUS video, containing a total of 466 images. Each example was manually segmented by expert clinicians to include predefined clinical features (sharp and fuzzy pleura, facsia bands, A lines, sub-A lines, vertical lines) on each frame. For each pixel within a segmented region, they assigned confidence scores reflecting their certainty (e.g., a value from 0 to 1) of that pixel representing each clinical feature.


Figure \ref{fig:confidence-label} is an example segmentation label, thresholded by confidence to show the distribution of confidences in a single example. As the confidence threshold increases, the resulting segmentation labels get more refined. However, there may be a loss of features in the labels once the thresholds get sufficiently higher than the expert's confidence in that feature's appearance. For example, vertical lines disappear once the threshold becomes 40\% in Figure \ref{fig:confidence-label}. By thresholding the confidence labels this way and conducting further analysis with downstream tasks, we can gain a sense of how prominent these LUS features need to be for them to be clinically relevant.

In addition to ultrasound videos, the dataset contains the following clinical information:
\begin{itemize}
    \item \textit{S/F Ratio}: The S/F represents the ratio between measured blood oxyhemoglobin saturation (S) and the fraction of inspired oxygen (F). The lower this ratio, the more deranged the lung function is. S/F is a standardized measurement used to assess lung function in research and at bedside. Each patient has one S/F ratio value per recorded day of their hospital stay, which includes their day of admission, the day after, and optionally, their day of discharge. In this dataset, this S/F ratio is normalized by dividing it by a maximum value of 477. 
    \item \textit{30-Day CHF Readmission}: The dataset contains a binary variable for each patient that indicates whether the patient was readmitted to the hospital within 30 days after the initial discharge date. The reason for readmission did not necessarily have to be related to CHF.
\end{itemize}
Importantly, no personally identifiable information from patients was used in this work. Patients were asked for consent in this usage of their data prior to their participation, in accordance with the IRB-approval.

For segmentation, the dataset of 42 CHF patients is split patient-wise into a training, validation, and testing set, consisting of 32, 6, and 4 patients, respectively. For our downstream tasks, we split the dataset into 6 folds, one of which is reserved as a held-out test set with the same 4 test patients as the segmentation test set, and the others randomly selected to be approximately equal in size. In this way, we perform cross-validation in our model trainings, allowing for more robust evaluation of our segmentation models for the downstream tasks.

\subsection{Tasks and Models}

\subsubsection{Segmentation}
We empirically chose to explore one main segmentation model to identify select features in ultrasound images: the Feature Pyramid Network (FPN) \cite{fpn}. It takes as input a single grayscale ultrasound image and produces a six-channel segmentation map, each channel being a binary mask for each of our desired features. 


FPN extracts multi-scale features by building a pyramid of feature maps at different resolutions and using pathway connections to pool information across resolutions. This allows it to segment features of various sizes, which is important as lung features like A lines and pleural lines differ in scale.

We aim to leverage confidence values in the following two ways.
We first convert expert confidence segmentations to binary segmentation maps using a \textit{confidence threshold} between 0\% and 100\%, where pixels greater than or equal to the threshold would correspond to positive pixels in the binary segmentation. Exploring different confidence thresholds allows us to determine how "sure" clinicians need to be in the structures they label in order for the labels to be clinically relevant to create segmentations and use them for downstream clinical tasks. In particular, we try the following thresholds: 100\%, 80\%, 60\%, 50\%, 40\%, 20\%, and 0\%. The model trained with the 0\% threshold is considered a baseline approach, as it is equivalent to not thresholding the confidences at all. We also consider performance relative to the 50\% model, as it's a naive approach to thresholding confidence values.
Furthermore, we focus the model's efforts on the most confident pixels during training by weighting the loss function by the expert confidence labels. Specifically, each pixel's loss is weighted by the corresponding expert-supplied confidence. The background pixels are weighted by the model's confidence threshold, other than for the 100\% and 0\% confidences; in which cases the background confidence was set to 0.8. This method was empirically chosen after exploring other weighting schemes. In this way, we penalize the model more for getting higher-confidence pixels incorrect, replicating the experts' uncertainty in lower-confidence pixels.\\


Our segmentation models are evaluated using the following metrics:
\begin{itemize}
    \item Intersection over Union (IoU): a measure of overlap between the predicted and ground truth segmentations. The ground truth segmentations are thresholded binary versions of the expert-labeled confidence maps.
    \item Weighted cross-entropy loss value: a measure of per-pixel accuracy, taking into account the importance of each pixel based on the ground truth confidence values. The labels are thresholded binary expert segmentations.

    \begin{equation}
    \mathcal{L}_{\text{weighted}} = -w \cdot \left[ y \cdot \log(\hat{y}) + (1 - y) \cdot \log(1 - \hat{y}) \right]
    \end{equation}

    \item Cross-entropy loss (unweighted and unthresholded): a measure of per-pixel accuracy. The labels are unthresholded raw expert confidences.
    \item Trimap loss: a measure of per-pixel accuracy that calculates a cross-entropy loss on only the pixels that are certainly background or foreground (i.e., have either a confidence value of 0\% or 100\%). In this way, we calculate how each model performs on a single ground truth, allowing us to better compare model performance for different confidence thresholds.
\end{itemize}

Finally, the clinical significance of these segmentation models and the effect of confidence thresholds are evaluated through their performance on various downstream tasks.

\subsubsection{Downstream Task 1: S/F Ratio Change Prediction}
This task allows us to evaluate our segmentation models in a situation where examples are directly compared to each other. In particular, we use the segmentations in addition to the LUS videos to predict whether one's S/F ratio is larger, smaller, or the same compared to another's. In doing so, our segmentations need to contain enough information to determine significant differences in features between different LUS videos.

We first create a paired dataset from our collection of CHF video data. Each pair of data contains two LUS video examples, with the constraint that both videos should be from the same view zone so that the model has ample opportunity to compare features across similar body regions between the pair. During training, we allowed pairs to have examples across patients and days. During validation and testing, we additionally constrain the video pairs to be from the same patient, as evaluating pairs between patients is not as clinically significant. We constructed pair labels by comparing their S/F ratios and converting them into three label classes: "Decrease" (first S/F > second), "Increase" (first S/F < second), and "Same" (S/F equal). This process resulted in approximately 20,000 train pairs and 300 validation pairs per data fold, with a total of 200 test pairs. 

We use a late fusion approach to evaluate each pair. Independently for each video in an example pair, we first segment out the clinical features by running our trained segmentation models on every video frame, then add the segmentations as additional channels to the original grayscale frames. We then use the Temporal Shift Module (TSM) video network \cite{tsm} with the MobileNet-v2 \cite{mobilenetv2} backbone to extract video features of the lung ultrasounds and segmentations. TSM aims to provide the benefits and competitive performance of a 3D CNN while enjoying the complexity of a 2D CNN. It infuses temporal information into every 2D CNN resnet block by shifting certain channels from the previous and next time frame. Refer to \cite{tsm} for more details. 

This process results in a set of video and segmentation features for each video. To combine these features, we subtract the first video's features from the second video's features to get a set of combined features. These are finally passed through a classification head, which consists of one multi-layer perceptron layer, to get the final prediction logits.

The model is trained using a cross-entropy loss and evaluated in two ways. First, we consider the classification accuracy of predicting a pair as "Decrease", "Same", or "Increase". Additionally, we combine the "Decrease" and "Same" classes to evaluate the models as a 2-class problem: "Increase" or "Not Increase". Because an increase in S/F ratio is associated with improvement in lung function, evaluating whether the S/F ratio increases for a certain patient across two days is clinically relevant to determine the patient's progress.

\subsubsection{Downstream Task 2: S/F Ratio Estimation}
This task aims to estimate the S/F ratio that a patient exhibited on a specific day of their hospital stay, requiring us to combine information from multiple views to get a final S/F ratio for the patient.

For each of the six views, we first independently predict an S/F ratio from its LUS video. We use the same TSM network \cite{tsm} on the LUS grayscale video and segmentations to extract video features. We add a regression head that consists of two multi-layer perceptron layers, each with 256 and 64 hidden nodes, respectively, and ReLU activations. The output of the regression head provides a single value that aims to represent the patient's S/F ratio.

Once we do this for all views, we explore ways to combine the answers from different views into a single, patient-level prediction. In particular, we consider taking an average, median, and max of individual S/F predictions.

The model (TSM and regression head) are trained using a Mean-Squared Error (MSE) loss. We evaluate performance in this task using the Root Mean Squared Error (RMSE) metric, where a lower value indicates better model performance.

\subsubsection{Downstream Task 3: 30-Day CHF Readmission}
Finally, we aim to predict whether a patient will be readmitted to the hospital within 30 days of their initial discharge date - a complex task involving all six LUS views over two time points to create a single "yes" or "no" prediction per patient.

We use the same approach as the other tasks to extract video features from video frames and segmentations using the TSM video network \cite{tsm}. However, we initialize the TSM using pretrained weights of the relevant S/F Ratio Change model (see task 2), allowing the network to learn LUS features easier and faster while adapting to the readmission task. This is done for each view and day separately, resulting in 12 feature vectors (2 days for each of the 6 views). Then the days are combined for each view by subtracting the first day's feature vector from the second day's feature vector, resulting in 6 combined feature vectors (1 per view). Each of them independently goes through a classification head consisting of a multi-layer perceptron layer to get the readmission prediction.

To combine predictions from different views, we take a majority vote (i.e., the mode) of the predicted readmission values. To break ties, we add the raw readmission logits over all views and consider the final answer to be the class that has the largest logit sum.

The model is trained using a cross-entropy loss and evaluated for accuracy, recall, and precision.

\subsection{Experiments}
All models were implemented in PyTorch and trained using Pytorch Lightning. All experiments were run using an NVIDIA RTX A6000 GPU. The total compute used included the experiments mentioned below as well as ablation studies for both the segmentation work and downstream tasks.

\subsubsection{Segmentation}
We trained separate models with different confidence thresholds: 100\%, 80\%, 60\%, 50\%, 40\%, 20\%, and 0\% (no threshold). To help mitigate the limitations of the small dataset size, data augmentation was used to create versions of the LUS images and labels: horizontal flips, rotations within 15 degrees, and an intensity transform were probabilistically applied randomly to images in the training set during every epoch of training. This allows the model to see more unique images, allowing for it to overfit less and be more robust. 

Each model was trained for 100 epochs using the Adam optimizer \cite{Kingma2015Adam:Optimization} with an initial learning rate of 0.0001, chosen empirically after experiments on a small portion of the dataset. A cosine annealing scheduler \cite{Loshchilov2016SGDR:Restarts} was used to alter the learning rate over the course of training. Each model took around 4.5-5.5 hours and 33 GB of GPU memory to train, depending on the loss function used. The highest average IoU over all the validation examples during training was used to select the best model, which was used to evaluate the test set examples.

We use a weighted cross-entropy loss to train our segmentation models, which was empirically chosen based on an ablation study we conducted with several other loss functions. The weighting scheme used is described in Section 2.2.1.



\subsubsection{Downstream Tasks}
Each trained segmentation model is frozen and used separately in each of the downstream tasks.

We use 5-fold cross validation with the same data splits for all downstream tasks to evaluate the robustness of our models. After training and validation, all models are evaluated on the same held-out test set of 4 patients. The variability of our results is reported using the standard deviation (1-sigma error bars) of the evaluation metrics across the 5 cross-validation data splits.

All models were trained using the Adam optimizer with an initial learning rate as either 0.0001 or 0.00001, chosen empirically from small experiments evaluated on the validation set for one cross-validation split. Cosine Annealing with warm restarts was used as a learning rate scheduler during training. Models took approximately 1.5 hours, 15 minutes, and 5 minutes for tasks 1, 2, and 3, respectively, making it relatively computationally efficient to train these models. Given the dataset size difference between Task 1 and the remaining tasks, these models scale relatively well with dataset size. Training these models took between 12 and 22 GB of GPU memory, with task 1 needing the most allocated memory.
 




\section{Results}

\subsection{Segmentation}
\begin{figure}[h]
  \centering
  \begin{minipage}{0.48\textwidth}
    \centering
     \includegraphics[width=1.0\linewidth]{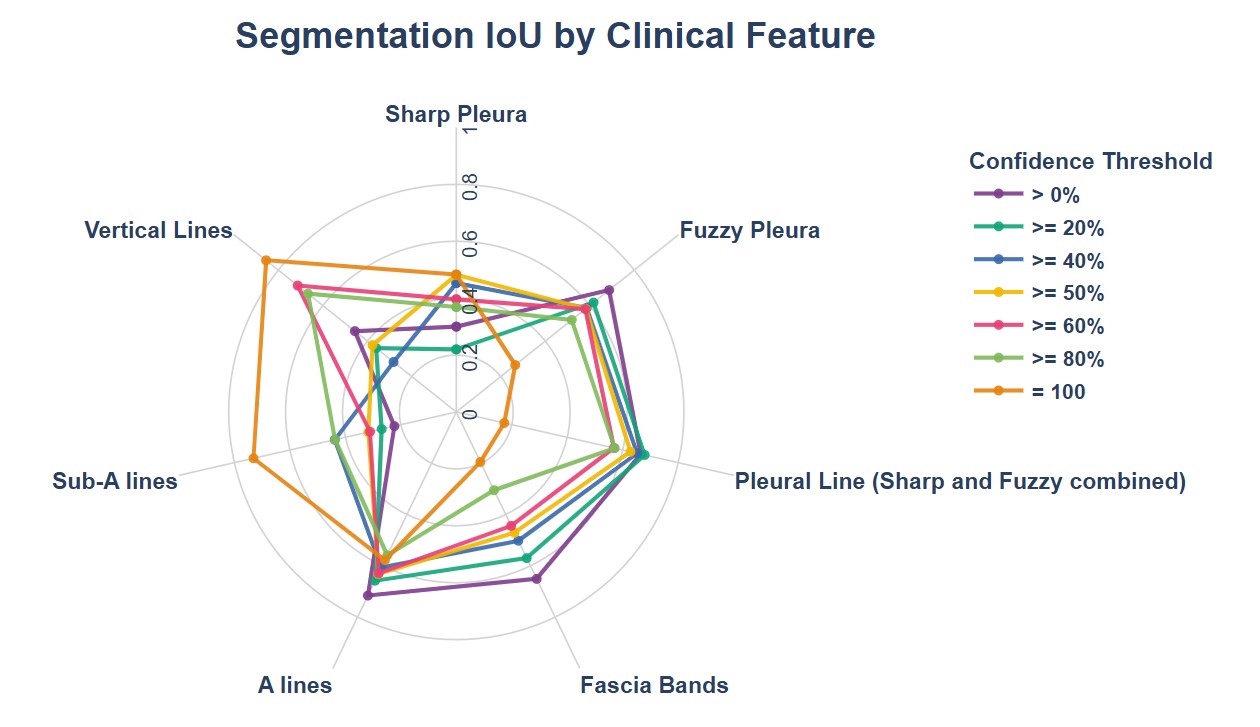}
    \caption{Segmentation model performance for various lung ultrasound structures.}
    \label{fig:seg-spider}
  \end{minipage}
  \hfill
  \begin{minipage}{0.48\textwidth}
\centering
  \captionof{table}{Segmentation model performance across confidence thresholds. We observe a disparity between the standard IoU metric and the three loss metrics, whose performance better reflects downstream clinical task outcomes.}
  \label{tab:seg-results}
 \resizebox{\columnwidth}{!}{
  \begin{tabular}{ccccc}
    \toprule
    \parbox{2cm}{\centering Confidence\\Threshold} &  
    IoU $\uparrow$ &                                              
    \parbox{1.8cm}{\centering WeightedCE\\Loss } $\downarrow$ & 
    \parbox{1.8cm}{\centering CE Loss\\(Unthresholded) } $\downarrow$ &
    \parbox{1.6cm}{\centering Trimap\\Loss} $\downarrow$ \\
    \midrule
    $> 0\%$ & \textbf{0.597} & 0.055 & 0.177 & 0.099 \\
    $\ge 20\%$ & \underline{0.535} & 0.022 & 0.075 & 0.067 \\
    $\ge 40\%$ & 0.478 & 0.022 & 0.053 & 0.039 \\
    $\ge 50\%$ & 0.424 & 0.024 & \underline{0.051} & 0.037 \\
    $\ge 60\%$ & 0.432 & \underline{0.018} & \textbf{0.050} & \underline{0.029} \\
    $\ge 80\%$ & 0.351 & 0.022 & 0.059 & 0.032 \\
    $= 100\%$ & 0.131     & \textbf{0.007} & 0.103 & \textbf{0.007} \\
    \bottomrule
  \end{tabular}
  }
  \end{minipage}
\end{figure}

The results of our segmentation models across the various confidence thresholds are summarized in Table \ref{tab:seg-results}. As expected, the IoU decreases as the confidence threshold increases, as segmentation maps from lower confidence thresholds tend to have larger foreground areas, increasing the chance of overlap with a predicted segmentation. The WeightedCE, CE (Unthresholded), and Trimap losses are more standardized metrics that allow for a closer comparison of segmentation performance across confidence thresholds. The 100\% threshold model performs the best on the Trimap loss, which was expected since this model optimizes for the 100\% confidence labels. Beyond the 100\% model, these more clinically aligned metrics demonstrate that models trained with higher confidence thresholds, particularly the 60\% model, show more promise in segmentation performance compared to the 0\% baseline naive 50\% model and the 0\% baseline. Notably, all models trained with any threshold (higher than 0\%) perform better in terms of the WeightedCE, the CE (unthresholded), and Trimap loss compared to the 0\% model, strongly suggesting that leveraging confidence labels by thresholding significantly improves segmentation accuracy on confidently labeled pixels.


The performance of our models in segmenting specific lung ultrasound features is represented in Figure \ref{fig:seg-spider}. None of the models demonstrated a consistent advantage across all features, highlighting that each model possessed relative strengths and weaknesses depending on the visual characteristics of the target. For example, the model trained with 100\% confidence labels performed best at segmenting features that typically appear as intensely bright, relatively isolated large areas, such as certain vertical artifacts (B-lines) and sub-A lines. These features may be easier to delineate due to their high contrast and limited overlap with surrounding structures. In contrast, models trained with lower thresholds, such as the 50\% confidence model, performed relatively well on pleural features, indicating that this model is potentially better at identifying structures that are less uniformly bright or more linear and contiguous, such as the pleural line itself, which requires recognizing a specific pattern along a boundary. Across the models, there was a general tendency to struggle with thin, elongated structures like a sharp pleural line and fine sub-A lines, while performing better on features that occupy larger areas of the image or have a more amorphous appearance, such as a thickened, "fuzzy" pleura or fascia bands in the chest wall.

\begin{figure}[h]
\centering
\begin{tabular}{@{}c@{\hspace{0.5em}}c@{\hspace{0.5em}}c@{\hspace{0.5em}}c@{\hspace{0.5em}}c@{\hspace{0.5em}}c@{\hspace{0.5em}}c@{\hspace{0.5em}}c@{}}
    LUS Image & $> 0$ & $\geq 20$ & $\geq 40$ & $\geq 50$ & $\geq 60$ & $\geq 80$ & $= 100$ \\
    \includegraphics[width=0.1\textwidth]{U5a1_image.jpg} &
    \includegraphics[width=0.1\textwidth]{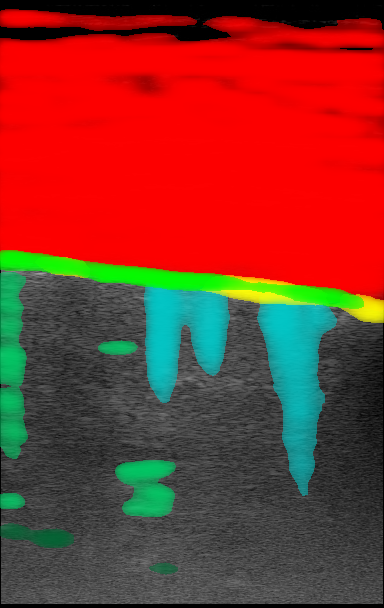} &
    \includegraphics[width=0.1\textwidth]{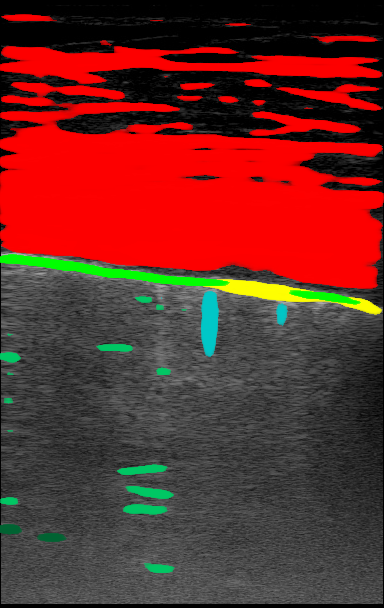} &
    \includegraphics[width=0.1\textwidth]{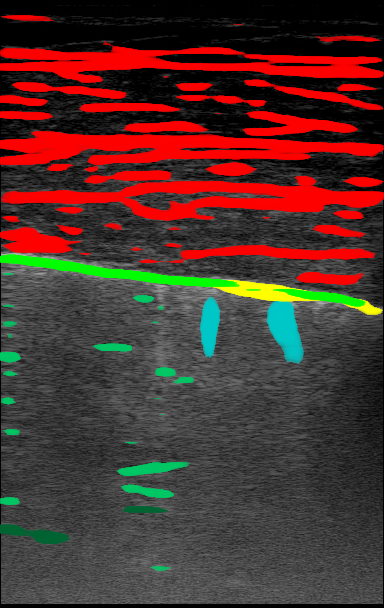} &
    \includegraphics[width=0.1\textwidth]{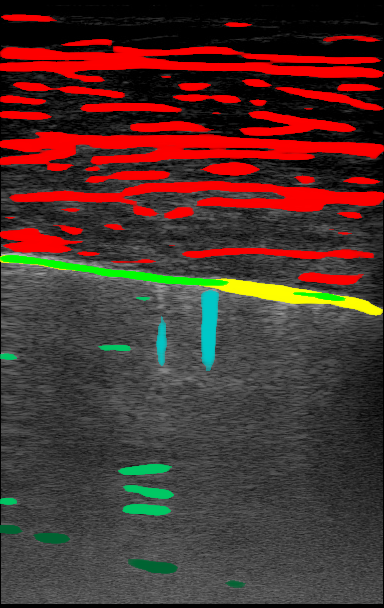} &
    \includegraphics[width=0.1\textwidth]{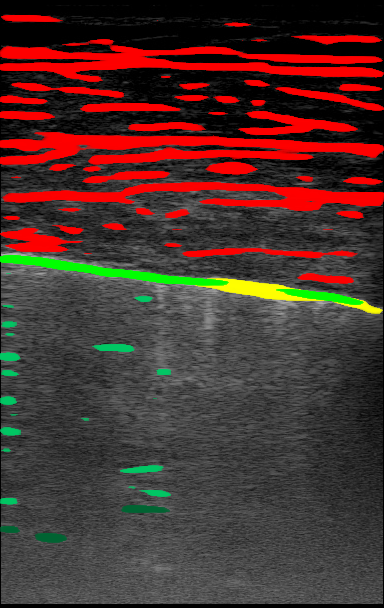} &
    \includegraphics[width=0.1\textwidth]{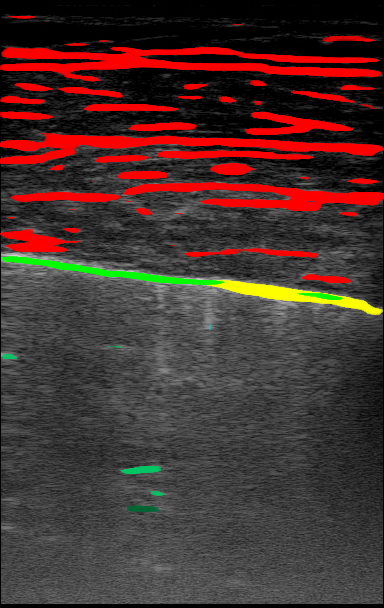} &
    \includegraphics[width=0.1\textwidth]{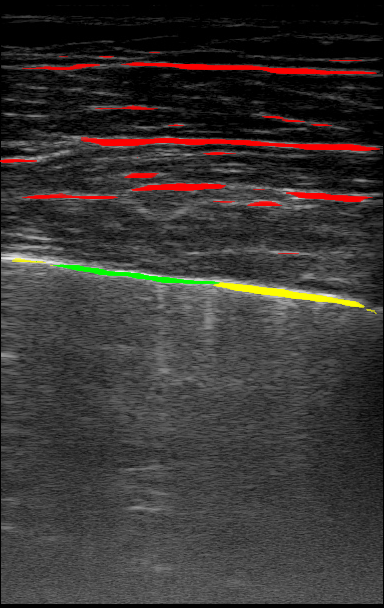}
\end{tabular}
\caption{Example segmentation outputs of models trained with various confidence thresholds.} 
\label{fig:pred-segmentations}
\end{figure}

Figure \ref{fig:pred-segmentations} shows example segmentation outputs for the same test example shown in Figure \ref{fig:confidence-label}. Each image represents the segmentation output of a model trained with the relevant confidence threshold. Similar to the expert labels, the segmented clinical features tend to get more refined as the confidence threshold increases. The 100\% model tends to over-segment features, while the 100\% model misses some features entirely, as expected.

\subsection{Downstream Tasks}
Evaluating our segmentation models on downstream tasks allows us to explore the balance between over-segmenting and under-segmenting clinical features in the LUS images, and, crucially, to determine if training on higher confidence labels translates to improved performance on clinically relevant predictions.


\begin{table}[h!]
  \centering
  \caption{S/F Change performance across various confidence thresholds.}
  \label{tab:results-sf-change}
   \begin{tabular}{@{}c*{2}{c}@{}} 
    \toprule
    {\parbox{2.3cm}{\centering Confidence\\Threshold}}
    & \parbox{2.3cm}{\centering 3-class \\ Accuracy $\uparrow$} 
    & \parbox{2.3cm}{\centering 2-class \\ Accuracy $\uparrow$} \\
    \midrule
    $> 0\%$    & 0.362 $\pm$ 0.033 & 0.497 $\pm$ 0.055 \\
    $\ge 20\%$  & 0.360 $\pm$ 0.049 & \textbf{0.531 $\pm$ 0.061} \\
    $\ge 40\%$  & \textbf{0.363 $\pm$ 0.031} & 0.515 $\pm$ 0.057 \\
    $\ge 50\%$  & 0.333 $\pm$ 0.022 & 0.501 $\pm$ 0.025 \\
    $\ge 60\%$  & 0.348 $\pm$ 0.033 & 0.515 $\pm$ 0.504 \\
    $\ge 80\%$  & 0.339 $\pm$ 0.049 & 0.504 $\pm$ 0.057 \\
    $\ge 100\%$ & 0.327 $\pm$ 0.051 & 0.511 $\pm$ 0.051 \\
    \bottomrule
  \end{tabular}
\end{table}

\subsubsection{S/F Change}
Table \ref{tab:results-sf-change} shows our results on the first downstream task, evaluated both as a 3-class and 2-class problem. While the overall performance of our models was only slightly above chance for this task, when comparing the impact of confidence thresholds, models trained with higher thresholds generally performed better than the 0\% baseline and the naive 50\% model on the more clinically relevant 2-class accuracy metric ("Increase" vs "Not Increase"). Although the 1-sigma error bars suggest that the differences were not statistically significant, there is a trend towards improved performance with increased confidence thresholds.

\begin{table}[h!]
  \centering
  \caption{S/F Prediction performance across various confidence thresholds.}
  \label{tab:results-sf-prediction}
   \begin{tabular}{@{}c*{3}{c}@{}} 
    \toprule
    {\parbox{2.3cm}{\centering Confidence\\Threshold}}
    & RMSE-Max $\downarrow$
    & RMSE-Average $\downarrow$
    & RMSE-Median $\downarrow$ \\    
    \midrule      
    $> 0\%$     & 0.3506 $\pm$ 0.0567 & 0.3964 $\pm$ 0.0568 & 0.3918 $\pm$ 0.0548 \\
    $\ge 20\%$  & 0.3708 $\pm$ 0.0580 & 0.4284 $\pm$ 0.0641 & 0.4190 $\pm$ 0.0641 \\
    $\ge 40\%$  & 0.3644 $\pm$ 0.0136 & 0.4188 $\pm$ 0.0112 & 0.4090 $\pm$ 0.0123 \\
    $\ge 50\%$  & 0.3644 $\pm$ 0.0245 & 0.4186 $\pm$ 0.0245 & 0.4078 $\pm$ 0.0225 \\
    $\ge 60\%$  & \textbf{0.2902 $\pm$ 0.0496} & \textbf{0.3604 $\pm$ 0.0533} & \textbf{0.3506 $\pm$ 0.0526} \\
    $\ge 80\%$  & 0.3468 $\pm$ 0.0439 & 0.4076 $\pm$ 0.0550 & 0.3948 $\pm$ 0.0531 \\
    $\ge 100\%$ & 0.3270 $\pm$ 0.0432 & 0.3822 $\pm$ 0.0362 & 0.3750 $\pm$ 0.0389 \\
    \bottomrule
  \end{tabular}
\end{table}

\subsubsection{S/F Prediction}
Table \ref{tab:results-sf-prediction} presents the performance of our models on the S/F Ratio Estimation task. Across all methods of combining view predictions (Max, Average, and Median), models trained with higher confidence thresholds consistently outperformed the 0\% baseline and the naive 50\% model. Notably, the model trained with the 60\% confidence threshold achieved the lowest RMSE across all aggregation methods, indicating that training on more confidently labeled pixels leads to more accurate S/F ratio estimations. This supports the claim that leveraging higher confidence labels improves performance on downstream tasks.

\begin{table}[h!]
  \centering
  \caption{CHF Readmission performance across various confidence thresholds. 
  }
  \label{tab:results-chf-readmission}
   \begin{tabular}{@{}c*{3}{c}@{}} 
    \toprule
    {\parbox{2.3cm}{\centering Confidence\\Threshold}}
    & \parbox{2.3cm}{\centering Accuracy} $\uparrow$
    & \parbox{2.3cm}{\centering Recall} $\uparrow$
    & \parbox{2.3cm}{\centering Precision} $\uparrow$ \\
    \midrule
    $> 0\%$     & 0.450 $\pm$ 0.209 & 0.400 $\pm$ 0.418 & 0.333 $\pm$ 0.312 \\
    $\ge 20\%$  & 0.550 $\pm$ 0.209 & 0.500 $\pm$ 0.354 & 0.567 $\pm$ 0.365 \\
    $\ge 40\%$  & 0.400 $\pm$ 0.137 & 0.600 $\pm$ 0.418 & 0.367 $\pm$ 0.217 \\
    $\ge 50\%$  & 0.550 $\pm$ 0.112 & 0.700 $\pm$ 0.274 & 0.533 $\pm$ 0.075 \\
    $\ge 60\%$  & \textbf{0.750 $\pm$ 0.177} & \textbf{0.900 $\pm$ 0.224} & \textbf{0.768 $\pm$ 0.224} \\
    $\ge 80\%$  & 0.700 $\pm$ 0.274 & 0.800 $\pm$ 0.274 & 0.700 $\pm$ 0.274 \\
    $\ge 100\%$ & 0.700 $\pm$ 0.209 & 0.800 $\pm$ 0.274 & 0.667 $\pm$ 0.204 \\
    \bottomrule
  \end{tabular}
\end{table}

\subsubsection{CHF Readmission}
Results for this task are shown in Table \ref{tab:results-chf-readmission}. We observe that all thresholded models outperform the 0\% baseline, with performance metrics generally showing an increasing trend as the confidence threshold increases, suggesting that including too many features in segmentations could be detrimental to determining readmission potential. In particular, the 60\% model outperforms the rest of our models in all three of our chosen metrics. Error bars are large for all models due to the small number of per-patient datapoints (4) in our test set. Further testing with a larger dataset will be needed to determine statistical significance of this model's results.

Overall, our downstream task analysis preliminarily shows that a 60\% confidence threshold is the most optimal threshold to guide our segmentation models for the best performance in clinical downstream tasks. We particularly note that this result loosely corresponds to segmentation performance as evaluated by the WeightedCE and Trimap metrics. More importantly, models trained with higher thresholds tend to perform better than the 0\% baseline model and the naive 50\% model in almost all of our tasks, indicating that leveraging confidence values in training segmentation models could lead to better downstream performance.

\section{Limitations and Future Work}
The major limitation of this work is the relatively small dataset size. While providing promising insights, the limited number of patients, particularly in downstream task test sets (only 4 for readmission), results in large error bars. This challenges definitive statistical significance and limits generalizability to a broader patient population. A larger, more diverse dataset would provide greater statistical power and allow for more robust validation of the proposed methodology. Furthermore, a larger dataset with more diverse data would be crucial for possible future translation to clinical application; failing to thoroughly incorporate and evaluate more diverse data could prevent generalizability of our methodology and unfairly impact certain patient populations.

Secondly, the expert-provided confidence values were not standardized across different clinicians. This introduces potential variability and inconsistency in the ground truth labels, as the interpretation and scoring of uncertainty may differ from one annotator to another. The lack of a standardized confidence annotation protocol means that the models are trained on a potentially heterogeneous representation of uncertainty. Future work could explore standardizing how clinicians assign confidence levels to address this in the labeling stage. Alternatively, if multiple annotators are given the opportunity to segment the same images, inter-annotator agreement could be incorporated into the methodology in using confidence values to address this in the analysis stage. 

These limitations underscore the need for larger-scale studies and the development of standardized annotation guidelines for confidence-aware labeling in medical imaging.

\section{Conclusion}

In this work, we introduced a novel methodology for leveraging expert-provided per-pixel confidence levels in training AI models for lung ultrasound segmentation, moving beyond the traditional approach of treating annotations as absolute ground truth. This method demonstrates that explicitly modeling the inherent uncertainty in subjective medical imaging modalities like lung ultrasound can lead to improved segmentation accuracy. More critically, this enhanced segmentation quality from models trained with higher confidence thresholds consistently translates to better performance on crucial downstream clinical tasks, including classifying S/F ratio change, estimating S/F ratio values, and predicting 30-day patient readmission, suggesting that leveraging label confidence can substantially boost the reliability and clinical utility of AI in medical imaging. Specifically, our findings indicate that a segmentation model trained with a 60\% confidence threshold achieved the highest diagnostic accuracy on these downstream tasks. Future work with larger, multi-site datasets and standardized confidence annotation protocols will be essential to confirm these findings and translate this approach into widespread clinical application.

\section{Acknowledgments}
This research was supported in part by the Liang Zhao Endowed Fellowship and the Center for Machine Learning and Health (CMLH) Translational Fellowship at Carnegie Mellon University (CMU). The authors gratefully acknowledge the clinicians at Louisiana State University (LSU) for their assistance with data collection, as well as our collaborators at CMU for their valuable insights. All data used in this study were de-identified prior to analysis. The private LSU dataset, collected under IRB protocol number 1509 titled Artificial Intelligence Interpretation of Lung Ultrasound Images, was de-identified before being transferred to CMU.

\bibliographystyle{plainnat}
\bibliography{references,references_old}


\end{document}